# Oscillation death in diffusively coupled oscillators by local repulsive link

C.R. Hens[1], Olasunkanmi I. Olusola[1,2], Pinaki Pal[3], Syamal K. Dana[1]

[1]CSIR-Indian Institute of Chemical Biology, Jadavpur, Kolkata 700032, India
[2]Department of Physics, Federal University of Agriculture, Abeokuta, Nigeria
[3]Department of Mathematics, National Institute of Technology, Durgapur 713209, India

A death of oscillation is reported in a network of coupled synchronized oscillators in presence of additional repulsive coupling. The repulsive link evolves as an averaging effect of mutual interaction between two neighboring oscillators due to a local fault and the number of repulsive links grows in time when the death scenario emerges. Analytical condition for oscillation death is derived for two coupled Landau-Stuart systems. Numerical results also confirm oscillation death in chaotic systems such as a Sprott system and the Rössler oscillator. We explore the effect in large networks of globally coupled oscillators and find that the number of repulsive links is always fewer than the size of the network.

A quenching or death of oscillation is an important phenomenon [1-3] in coupled oscillators (limit cycle or chaotic) besides synchronization [4]. It is mainly dictated by large parameter mismatch in coupled oscillators [5] or delay in coupling [6] of identical oscillators. In recent times, several other mechanisms of oscillation death or stabilization of fixed point were reported using different coupling schemes which were based on dynamic coupling [7], mean field diffusion coupling [8, 9] and conjugate coupling [10] in identical oscillators, and dynamic environment coupling [11] in identical or mismatched oscillators. Of particular interest is the dynamic environment coupling [11] that is able to induce oscillation death in a network [12], chain, ring, tree, lattice, all-to-all, star, and random topologies. An over-damped dynamic environment influences each of the dynamical units in a network and suppresses the oscillation of all the units for a critical coupling.

In real world, a different situation may arise when besides the diffusive attractive coupling between the dynamical nodes that establishes *a priori* synchrony in a network of oscillators, additional coupling links or bonds evolve in time between two neighboring nodes in the network due to a local disturbance or a fault. This local fault can act as a repulsive feedback link on an immediate local node. We assume that the number of repulsive links increases in time to spread into the other nodes of the network. Eventually, the increasing repulsive links influence the dynamics of the network in time and induce a death situation as quenching of oscillation much before it spreads into the whole network. The concept of all-to-all additional dynamic environment coupling or links [12] cannot explain such a situation since only a fewer nodes than the size of the network are locally affected by the additional repulsive links and suffice to induce a death. We mention that a quenching of oscillation, although in a different context but of similar effect, was reported earlier as an aging transition [13] when, in a network of diffusively coupled oscillators, individual oscillators switch over to a passive state or excitable state one after another in time and that the oscillation in the network eventually comes to a stop when a sufficient number of oscillators switches over to the passive phase. Instead, we propose that the repulsive feedback links spread into *a priori* synchronized network attacking one after another oscillator and stop the oscillation of the network. To model this situation, we propose a coupling scheme with additional local repulsive links in Fig. 1(a) for simply two identical oscillators first: oscillator (2) drives the oscillator (1) attractively but, in addition, a local link returns to the oscillator (2) as a negative feedback called as the repulsive link. Alternatively, we can consider two oscillators under bidirectional attractive coupling as shown in Fig. 1(b) when the repulsive link may return to either of the oscillators. The repulsive link is expressed as an average of the state variables, $0.5\varepsilon_2(x_1 + y_1)$ of two neighboring dynamical units, $\dot{x}=f(x)$ and $\dot{y}=f(y)$ ; $x \in R^n$, $y \in R^n$ where the coupling strengths $\varepsilon_1$ is positive real and $\varepsilon_2$ is negative real. An average effect in terms of a negative feedback from local oscillators similar to the repulsive link is used earlier [14] in globally coupled network of oscillators in the context of deep brain stimulation (DBS) for control of neurological diseases where the robustness of a synchronized state or desynchronization was the target of the study. Later this average feedback effect is used [15] in globally coupled electronic oscillators for inducing a transition from synchronization to desynchronization. However, in both the cases, all the oscillators of the networks were assumed directly influenced by the feedback links in addition to the attractive coupling and no death scenario was reported.

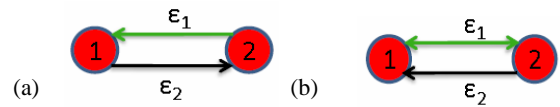

**Fig.1** (color online). Dynamical systems in circles (red) under attractive and repulsive coupling. Grey arrow (green) indicates unidirectional (a), bidirectional (b) attractive coupling between a pair of variables of the systems. Black arrows indicate repulsive link via the same or another pair of variables.

On the contrary, we report here a death scenario as a generic feature of oscillators (limit cycle and chaotic) when attractive diffusive coupling or link is mixed with local repulsive link. All the oscillators do not receive a repulsive feedback. Oscillators are assumed *a priori* synchronized under the unidirectional or bidirectional attractive coupling above a critical value $\varepsilon_1 > \varepsilon_{1C}$ ($\varepsilon_2 = 0$) before the repulsive link is added. In the absence of attractive diffusive coupling, the repulsive link induces out-of-phase in the oscillators for another critical value $\varepsilon_2 > \varepsilon_{2C}$ ($\varepsilon_1 = 0$). A mixture of both the coupling brings a competition and finally stops the oscillation to death which we investigate, in this brief report, with examples of two limit cycle and chaotic oscillators and a globally coupled network of oscillators.

Firstly, we investigate two coupled oscillators (limit cycle and chaotic) under both attractive and repulsive coupling where the oscillators emerge into a homogeneous steady state (HSS) or amplitude death (AD) [8] above a critical coupling. Interestingly, we find a transition from AD to inhomogeneous steady state (IHSS) state or oscillation death (OD) state with increasing coupling strength in both limit cycle and chaotic oscillators. Secondly, we study networks of N globally coupled synchronized oscillators and go on adding repulsive links one after another and address a relevant question how many repulsive links suffice to induce the death of oscillation (presently we search for AD only)?

We elaborate the repulsive coupling and the death scenario using two Landau-Stuart (LS) limit cycle systems,

$$\dot{x}_1 = [1-(x_1^2+y_1^2)]x_1 - \omega y_1 + \varepsilon_1(x_2 - x_1),$$
$$\dot{y}_1 = [1-(x_1^2+y_1^2)]y_1 + \omega x_1,$$
$$\dot{x}_2 = [1-(x_2^2+y_2^2)]x_1 - \omega y_2,$$
$$\dot{y}_2 = [1-(x_2^2+y_2^2)]y_2 + \omega x_2 - \varepsilon_2(y_1 + y_2).$$
(1)

where $\varepsilon_{1,2}$ is the coupling strength. Assume that the attractive coupling is applied in unidirectional mode to one oscillator via the *x*-variable while the repulsive link is returned to the other oscillator as a negative feedback via the *y*-variable. We derive the analytical conditions for AD in two LS systems. For simplification, we consider the case of symmetric coupling, $\varepsilon_1=\varepsilon_2=\varepsilon$ when the stability condition for the trivial equilibrium origin of the unidirectionally coupled LS system (1) is derived from its *Jacobian*,

$$J = \begin{bmatrix} 1-\varepsilon & -\omega & \varepsilon & 0 \\ \omega & 1 & 0 & 0 \\ 0 & 0 & 1 & -\omega \\ 0 & -\varepsilon & \omega & 1-\varepsilon \end{bmatrix}$$

whose eigenvalues are,

$$\lambda_{1,2} = \frac{-(\varepsilon-2) \pm \sqrt{(\varepsilon-2)^2 - 4\{1-\varepsilon(1+\omega)+\omega^2\}}}{2}$$

and $\lambda_{3,4} = \frac{-(\varepsilon-2) \pm \sqrt{(\varepsilon-2)^2 - 4\{1-\varepsilon(1-\omega)+\omega^2\}}}{2}$

This provides a stability condition of the equilibrium origin,

$$2 < \varepsilon < (\omega^2 + 1)/(\omega + 1) \qquad (2)$$

As an example, AD can be observed in two LS systems for a coupling range, $2<\varepsilon<2.5$ when $\omega=3.0$. For bidirectional attractive coupling with one repulsive link, the *Jacobian* can also be derived when the eigenvalues are,

$$\lambda_{1,2} = -(\varepsilon-1) \pm \sqrt{\varepsilon^2 - \omega^2}$$

and $\lambda_{3,4} = \frac{-(\varepsilon-2) \pm \sqrt{(\varepsilon-2)^2 - 4(1+\omega^2-\varepsilon)}}{2}$

when the stability of the equilibrium origin is decided by,

$$2 < \varepsilon < (\omega^2 + 1)/2 \qquad (3)$$

Clearly, the range of coupling (AD regime) increases with frequency $\omega$ for unidirectional as well as bidirectional attractive coupling similar to the case of delay coupling [6, 16]. As mentioned above, the coupled oscillators are assumed synchronized before adding the repulsive link in both the cases, however, details are not presented here.

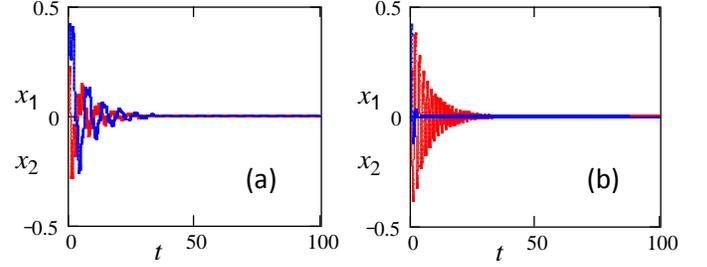

**Fig. 2** (color online). Amplitude death in two coupled LS systems with one repulsive link. Numerical time series of $x_1(t)$ and $x_2(t)$ show damped oscillation to origin for attractive (a) unidirectional, (b) bidirectional coupling. $\omega=3.0$, $\varepsilon_1=\varepsilon_2=\varepsilon=2.25$.

Numerical results of two coupled LS systems are shown in Fig. 2 using the system parameters and the coupling based on the condition (2). The time series of $x_1$– and $x_2$–variables in Fig. 2(a) show AD when the equilibrium origin is stabilized. Similarly, for bidirectional attractive coupling, the AD scenario is obtained following the stability condition (3) as shown in Fig. 2(b). For other parameter regimes, OD (IHSS) [8] is observed. A complete scenario of AD and OD in two coupled LS systems is illustrated in bifurcation diagrams in Fig. 3 (using software package MATCONT [17]). Figure 3 shows bifurcations of the coupled LS systems for unidirectional as well as bidirectional attractive coupling via *x*-variable and one repulsive link via *y*-variable. For a choice of $\omega=3.0$, an AD window in solid black line is obtained that emerges as a reverse-Hopf bifurcation with increasing coupling strength $\varepsilon$. This is indicated, in both the cases, by an existence of stable limit cycle (in grey line) that coexists with an unstable equilibrium origin (in black dashed line). On the right of the AD window, an OD state with two IHSS (in grey line) originates via supercritical pitchfork bifurcation (PB) and they coexist with an unstable equilibrium origin (in dashed black line).

Next, we search for a similar death scenario in chaotic oscillators under attractive diffusive coupling and repulsive link. We use two examples, a Sprott system and the Rössler system. Two coupled Sprott systems with attractive unidirectional coupling and a repulsive link are given by,

$$\dot{x}_1 = -ay_1, \quad \dot{y}_1 = x_1 + z_1 - \varepsilon(y_1 + y_2),$$
$$\dot{z}_1 = x_1 + y_1^2 - z_1$$
$$\dot{x}_2 = -ay_2 + \varepsilon(x_1 - x_2), \quad \dot{y}_2 = x_2 + z_2,$$
$$\dot{z}_2 = x_2 + y_2^2 - z_2$$
(4)

Numerical results show stabilization of the equilibrium origin in two Sprott systems in Fig. 4(a) where time series $x_1(t)$ and $x_2(t)$ converges to zero amplitude in time. The Sprott system has a trivial equilibrium point at origin. It clearly confirms an AD scenario in the coupled system for a coupling strength $\varepsilon=0.20$. In contrast, an OD scenario is observed in two coupled Rössler systems ($\dot{x}=-\omega y-z$, $\dot{y}=\omega x+ay$, $\dot{z}=b+z(x-c)$) in a chaotic regime where the time series converges to a stable fixed point in close proximity to the origin as shown in Fig. 4(b).

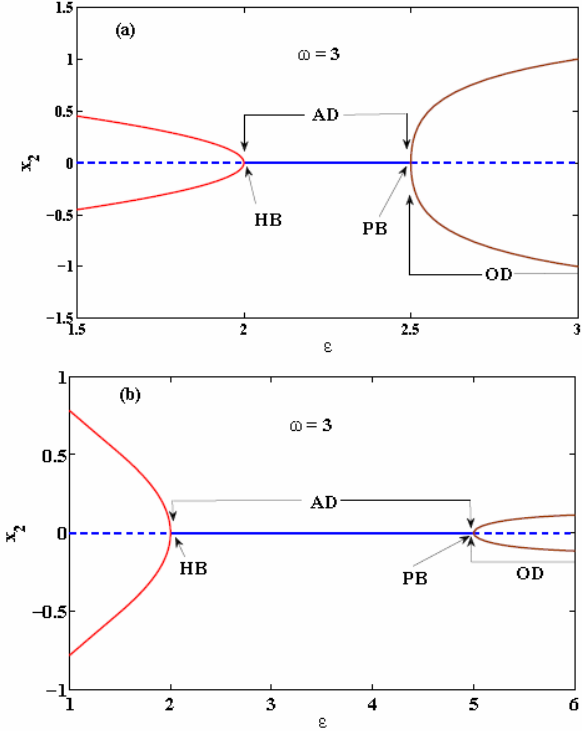

Fig. 3 (color online). Bifurcation diagram shows AD and OD regimes in coupled LS system. Values of $x_2$ for fixed point solutions and extremum of $x_2$ for time dependent solutions are plotted for (a) unidirectional diffusive coupling, (b) bidirectional diffusive coupling. Arrows indicate HB and PB points (using MATCONT software package [17]).

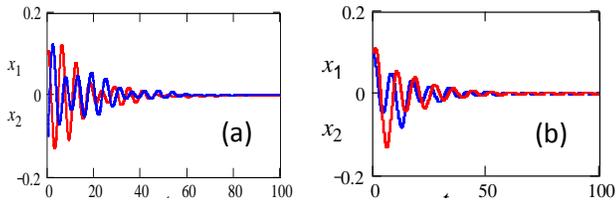

**Fig. 4** (color online). Death in coupled chaotic oscillators with repulsive link. Numerical time series of $x_1(t)$ and $x_2(t)$ plotted after initial transients. Equilibrium point is stabilized in (a) Rössler systems for $\omega=1.0$, $\varepsilon=0.225$, $c=18$, $b=a=0.1$, (b) Sprott systems for $a=0.225$, $\varepsilon=0.20$.

The analytical conditions of AD/OD in chaotic oscillators is difficult to derive and hence we draw a bifurcation diagram (using software package MATCONT [17]) in Fig. 5 to show the AD and OD regimes with varying coupling strength for two unidirectionally coupled chaotic Sprott systems with one repulsive link (for the model in eq. 4). Figure 5 clearly reveals an AD window (solid black line) of coupling strength, however, it coexists with an unstable equilibrium (grey dashed line). An OD (solid grey line) originates on the right hand side via transcritical bifurcation (TB) instead of PF as observed for coupled LS system and coexists with an unstable origin (grey dashed line). For lower coupling at left, two Hopf bifurcation points HB1 and HB2 appear related to stable (grey line) and unstable limit cycle (dashed black line) respectively.

Finally, we consider a population of N globally and diffusively coupled oscillators. We separate the network into two subpopulations, $p$ and $q$, such that $q$=N-$p$ oscillators are only connected by additional repulsive links, other are not.

Subpopulation 1:

$$\dot{x}_k = f(x_k) + \frac{\varepsilon}{N} \sum_{j=1}^{N}(x_j - x_k) \quad (5)$$

where
$$k = 1 \text{ to } p$$

Subpopulation 2:

$$\dot{x}_q = f(x_q) + [\frac{\varepsilon}{N} \sum_{j=1}^{N}(x_j - x_q)] - \varepsilon(x_k + x_q) \quad (6)$$

where
$$q = p+1 \text{ to } N$$

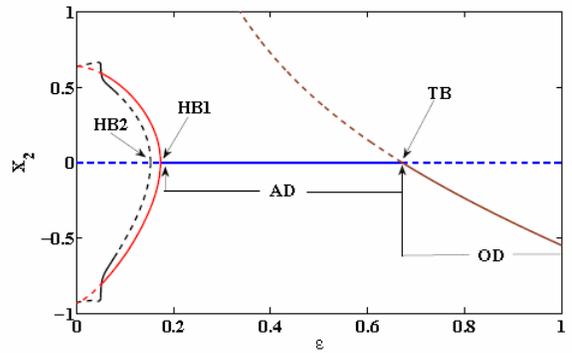

**Fig. 5** (color online). Bifurcation of AD and OD in two coupled Sprott systems, ($a$=0.225). Plotted values of $x_2$ for fixed point solutions and extremum values of $x_2$ for time dependent solutions of the coupled Sprott systems as a function of $\varepsilon$ (using MATCONT [17]). An AD regime in solid black (solid blue line) seen in the middle which changes to unstable equilibrium at a TB point. An unstable equilibrium coexists with the stable origin in grey dashed line (brown dashed) and switch over to OD at the TB point. At HB1 point a stable limit cycle in grey (solid red line) originates, an unstable limit cycle in black dashed line originates at HB2 point and coexists with unstable origin dashed black (dashed blue line).

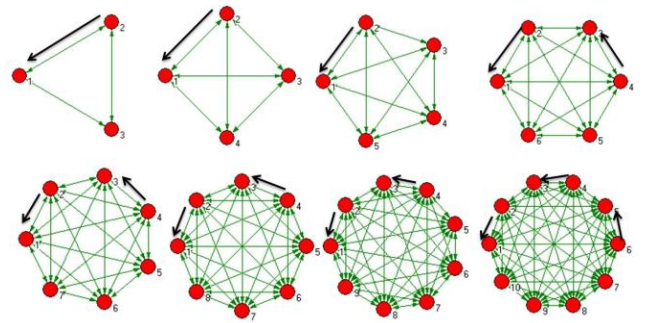

**Fig. 6** (color online). Networks of globally coupled identical LS oscillators. A circle (red) represents a LS oscillator in all eight networks. Black arrows indicate repulsive links and attractive coupling by grey (green) arrows. (Plotted using Pajek software [18]).

In such a population of N diffusively coupled synchronized oscillators, we start adding repulsive links to one after another oscillator and check the dynamics of the whole network until we add the links in *q*-oscillators. For an optimal number of repulsive links or *q*-oscillators connected with repulsive links, a death situation emerges. It is found that two populations or clusters of oscillators are formed which are synchronized separately with increasing coupling strength before reaching a death state which is basically an AD state for our example system.

The death scenario is elaborated with examples of eight networks of globally coupled N-oscillators (3, 4, 5, …. 10) as shown in Fig. 6. Each node in circle is a LS system in all the cases. It is confirmed from numerical simulations that one repulsive link (in black arrows) suffices to induce death in a network of 5-oscillators. Two repulsive links are necessary for a network of 9-oscillators and three for 10-oscillators.

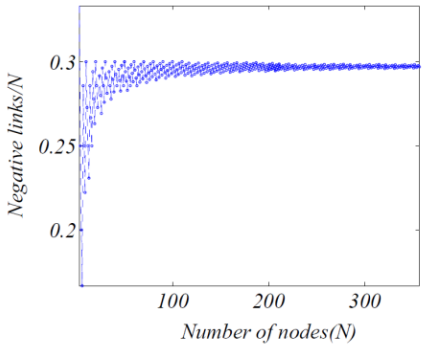

**Fig. 7** (color online). Globally coupled LS oscillators with repulsive links: N=10-350, ω=5.0, $\varepsilon_1=\varepsilon_2=8.0$. Plot of the percentage of repulsive links for inducing AD with size of the network (N).

Finally, we deal with the question what is the minimal number of repulsive links necessary for quenching oscillation in a large network of globally coupled oscillators? In a network of *a priori* synchronized oscillators under attractive coupling, the number of repulsive link increases with the size of the network to induce a death but a fewer repulsive links than the size of the network suffice to induce death in the network as described above. All the oscillators are coupled by attractive diffusive links whose number is always larger than the number of repulsive links for inducing a death of oscillation in the network. We produce numerical evidence (using Matlab ODE 45 solver) regarding the number of repulsive links necessary for inducing a death state in the networks. We simulate the dynamics of each globally coupled network of LS oscillators (N=10 to 350). To characterize the death or cessation of oscillation in each network, we use an index [13],

$$M = \sqrt{\left\langle (X_c - \langle X_c \rangle)^2 \right\rangle} \quad (7)$$

where $X_c = N^{-1}\sum_{j=1}^{N}(x_j, y_j)$ is the centroid for a network of LS oscillators. The bracket represents a time average when M=0 estimates a death scenario. For a network of size N (say, N=10), we keep adding repulsive links to one after another oscillator and simulate the collective dynamics to find the M-value and stop adding the repulsive link when it becomes zero. We repeat this protocol for all networks (N=10 to 350) to record the number of repulsive links until M=0 for each network. Figure 7 shows that the number of repulsive links certainly increases with N but it saturates around 30% of the total number of oscillators or the size of the network. In other sense, if a fewer number of nodes is directly influenced by the repulsive feedback, death of oscillation shall emerge for a critical coupling strength.

It is concluded that in a network of synchronized oscillators under diffusive attractive coupling, additional repulsive coupling links may evolve due to local fault and spread into the network by increasing number that eventually induce death of oscillation in the network. The number of repulsive links is always fewer than the number of oscillators or size of the network for a death scenario to emerge. We presented analytical conditions for AD in two coupled Landau-Stuart oscillators and numerical results for two coupled chaotic oscillators. Most interestingly, in the case of two oscillators, we found a transition from AD to OD with coupling strength via pitchfork bifurcation in Landau-Stuart system similar to what is reported [19] very recently although for a different coupling configuration. For two chaotic oscillators, we observed a similar AD to OD transition with coupling strength but following a different route, namely, a transcritical bifurcation which is new information, to our best knowledge. An analytical condition to find the critical number of repulsive links necessary for a death scenario in a large network is our future research target.

C.R.H. is supported by the CSIR Network project GENESIS (No.BC0121). O.I.O. is supported by the FICCI, India and the DST, India. S.K.D. is supported by the CSIR Emeritus scientist scheme. The authors appreciate Abhijit Sen and Jürgen Kurths for their valuable suggestions.
_______________________________________________